\documentclass[11pt]{article}
\usepackage{amsfonts}
\usepackage{amssymb, amsmath, cite}

\newtheorem{theorem}{Theorem}

\newtheorem{remark}{Remark}

\newcommand\be{\begin{equation}}
\newcommand\ee{\end{equation}}
\newcommand\ber{\begin{eqnarray}}
\newcommand\eer{\end{eqnarray}}
\newcommand\berr{\begin{eqnarray*}}
\newcommand\eerr{\end{eqnarray*}}
\newcommand{\ud}{\mathrm{d}}
\newcommand{\nm}{\nonumber}

\newcommand{\ito}{\int_{\Omega}}

\newcommand{\vep}{\varepsilon}
 \newcommand\wot  {W^{1, 2}(\mathbb{R}^2)}
\newcommand{\itr}{\int_{\mathbb{R}^2}}
\newcommand\re{\mathrm{e}}
{\setlength\arraycolsep{2pt}
\begin{document}
\begin{center}
 { \bf\Large  A Sharp Existence Theorem for Vortices in the Theory  of Branes\footnote{This work was supported by the Natural Science Foundation of China under grant  11201118.}}\\[2mm]
{Xiaosen Han\footnote{Email: \, xiaosenhan@gmail.com.} \\[2mm]
  {  Institute of Contemporary Mathematics\\
  School of Mathematics\\ Henan University\\ Kaifeng, 475004, PR China}}
\end{center}
\setlength{\baselineskip}{17pt}{\setlength\arraycolsep{2pt}

\begin{quote}{
{{\bfseries Abstract.}  We investigate the BPS equations arising from the theory of multiply intersecting    D-branes.  By using the
direct minimization method, we establish a sharp existence and uniqueness theorem  for multiple vortex solutions of the BPS
equations over a doubly periodic domain and over the full plane, respectively.  In particular, we obtain an explicit necessary and
sufficient condition for the existence  of a unique solution for the doubly periodic domain case.}}

{\bf Mathematics Subject Classification (2000).} 35A05, 58E50.

{\bf Keywords:}  Vortices; D-branes; BPS equations; nonlinear  elliptic system; direct minimization.

\end{quote}
\section{Introduction}\label{S:1}

Vortices play  important roles in many areas of theoretical physics including superconductivity theory \cite{abri,jata,gila},
condensed-matter physics \cite{igrc,kaoh}, optics \cite{bec}, and cosmology \cite{hiki,kibb,vish}. It is Taubes who first obtained the
rigorous construction of  multiple vortex static solutions for the Abelian Higgs model in  \cite{jata,taub1,taub2}. Since then a great
deal  of work has been done on various vortex equations. See  \cite{caya1,chsp,han1,hahu,haja,haki,hokp,jawe,rita,tara1,spya1,spya2,wangr,yang1}.
Motivated by the seminal work of Seiberg and Witten \cite{sewi} on monopoles condensation and confinement, followed by Hanany and Tong \cite{hato},
 considerable attention has been devoted to the studies on non-Abelian multiple vortices in supersymmetric gauge field theories
\cite{aubo,auku,efnn,eino1,eino2,gjk,shyu1,shyu2,shyu3}. For the rigorous existence of such vortices, Lieb and Yang \cite{liya}, Lin and Yang \cite{linyang1,linyang2} developed a
series of existence and uniqueness theories. Han and Tarantello \cite{hata1} established  existence of doubly periodic non-Abelian  Chern-Simons   vortices with a general gauge group.
 For more related work and references we refer to the monograph \cite{yang1}.

The purpose of this paper is to establish the existence of multiple vortex solutions  for the BPS equations derived in  \cite{suy2}  from
the theory of multi-intersection of D-branes. By complexifying the variables, we can formulate  the corresponding BPS equations into an
$l\times l\, (l\ge2)$ system of nonlinear elliptic equations. Then, using the direct minimization method developed in  \cite{liya}, we
can establish the existence and uniqueness of solutions to the BPS equations over a doubly periodic domain $\Omega$ and over the full
plane $\mathbb{R}^2$, respectively. It is worth noting  that we can establish an explicit necessary and sufficient condition for the
existence  of a unique solution for the doubly periodic domain case and such result is very rare in the existing literature.

The rest of our paper is organized as follows. In section 2, following Suyama \cite{suy2,suy3} we derive the BPS equation and
state our main results on existence and uniqueness of multiple vortex solutions for the BPS equations. Section 3 is devoted to  the
proof of the existence result in the doubly periodic domain case. In section 4 we prove the existence result for the planar case. In
section 5 we summarize our results and draw a conclusion.

\section{Vortices in the theory of Branes}\label{S:2}

\setcounter{equation}{0}

To formulate our problem, we follow Suyama  \cite{suy2,suy3}. We consider the following brane configuration in Type IIA theory
compactified on $T^6$. There are $Q_1$ D4-branes and $Q_2$ D4-branes wrapped along different directions of $T^6$. They intersect over a
3-dimensional hyperplane. The low energy effective theory on the D4-D4' intersection is a 3-dimensional gauge theory,
whose gauge group is $U(Q_1)\times U(Q_2)$. For simplicity, we assume that D4(D4')-branes are separated from each other.
Then the gauge group is reduced to $U(1)^{Q_1}\times U(1)^{Q_2}$. The action is the following,
 \ber
  S&=&\int\ud^3x\left[\sum\limits_{n=1}^{Q_1}\left(-\frac{1}{4g^2}F^{(1)}_{nij}F_n^{(1)ij}-\frac{g^2}{2}(D_n^{(1)})^2-g^2|F_n^{(1)}|^2\right)
 \right.\nonumber\\
 &&\left.+\sum\limits_{m=1}^{Q_2}\left(-\frac{1}{4g^2}F^{(2)}_{mij}F_m^{(2)ij}-\frac{g^2}{2}(D_m^{(2)})^2-g^2|F_m^{(2)}|^2
  \right)\right.\nonumber\\
  &&\left.-\sum\limits_{n=1}^{Q_1}\sum\limits_{m=1}^{Q_2}|D_iq_{nm}|^2\right],
  \,\,\, (i, j=0, 1, 2),\label{2.1}
 \eer
 where
   \berr
   D_iq_{nm}&=&\partial_iq_{nm}+\mathrm{i}\left(A^{(1)}_{ni}-A^{(2)}_{mi}\right)q_{nm},\\
   D_n^{(1)}&=&-\sum\limits_{m=1}^{Q_2}\big(|q_{nm}|^2-|\tilde{q}_{nm}|^2-\zeta\big),\\
    D_m^{(2)}&=&+\sum\limits_{n=1}^{Q_1}\big(|q_{nm}|^2-|\tilde{q}_{nm}|^2-\zeta\big),\\
    F_n^{(1)}&=&-\sqrt{2}\sum\limits_{m=1}^{Q_2}q_{nm}\tilde{q}_{mn}, \\
    F_m^{(2)}&=&+\sqrt{2}\sum\limits_{n=1}^{Q_1}\tilde{q}_{mn}q_{nm},
   \eerr
 $A^{(1)}_{ni}$ and $A_{mi}^{(2)}$ are the gauge fields in the Cartan subalgebra of $U(Q_1)$ and $U(Q_2)$.

Assume that $\zeta$ is positive, without potential we have the vacuum configuration
 \ber
  |q_{nm}|=\zeta, \,\,\, \tilde{q}_{mn=0}. \label{2.2}
 \eer
The vortex solution must satisfy this condition at spatial infinity.

 When $Q_1=l \,(\l\ge2),  Q_2=1$,  by the Bogomol'nyi reduction  \cite{bogo,jata} for the static solutions, Suyama  \cite{suy2} obtained the
 BPS equations  of the following  form,
 \ber
  &&F^{(1)}_{j,12}\pm g^2(|q_{j1}|^2-\zeta)=0,\quad j=1,\dots, l,\label{2.3}\\
  &&F_{1, 12}^{(2)}\mp g^2\sum\limits_{i=1}^l(|q_{i1}|^2-\zeta)=0,\label{2.4}\\
  &&\left[\partial_1q_{j1}+\mathrm{i}\left(A_{j,1}^{(1)}-A^{(2)}_{1, 1}\right)q_{j1}\right]
 \pm\mathrm{i}\left[\partial_2q_{j1}+\mathrm{i}\left(A_{j, 2}^{(1)}-A^{(2)}_{1, 2}\right)q_{j1}\right]=0,\quad j=1,\dots, l.\label{2.5}
  \eer

 Redefining the fields as follows
 \ber
  A_{j_i}=A_{j, i}^{(1)}-A_{1, i}^{(2)},
  \,\,i=1,2,\,\,
  F_j=\partial_1A_{j_2}-\partial_2A_{j_1},\,\, j=1,\dots,l \label{2.6}
 \eer
and using suitable re-scaling, the BPS equations \eqref{2.3}--\eqref{2.5} are transformed into
 \ber
  &&F_j+|q_{j1}|^2+\sum\limits_{i=1}^l|q_{i1}|^2-(l+1)=0,\,\,\,\,\,j=1,\dots,l,\label{2.7}\\
   &&(\partial_1q_{j1}+\mathrm{i}A_{j_1}q_{j1})-\mathrm{i}(\partial_2q_{j1}+\mathrm{i}A_{j_2}q_{j1})=0,\,\,\,\,\,j=1,\dots,l,\label{2.8}
 \eer
where we take lower sign in \eqref{2.3}--\eqref{2.5}.
As in  \cite{jata},  we can see from the equation \eqref{2.5} that the zeros of $q_{11},
\dots, q_{l1}$ are discrete and of integer multiplicities. Now we  denote the zero set  of $q_{j1}$ by $Z_{q_{j1}}$,
 \ber
 Z_{q_{j1}}=\{p_{j,1}, \dots, p_{j, N_j}\}, \quad\,\, j=1, \dots, l\label{2.10}
 \eer
such that the repetitions among the points take account of the  multiplicities of these zeroes.

 For the  equations \eqref{2.7}--\eqref{2.8}, we are interested in two cases. In the first case we study the equations   over a doubly periodic domain $\Omega$, governing
multiple vortices hosted in $\Omega$ such that the field configurations are subject to the 't Hooft boundary condition
 \cite{hoof,waya,yang1} under which periodicity is achieved modulo gauge transformations. In the second case we consider the equations
over the full plane $\mathbb{R}^2$  with the natural  boundary condition
\ber
|q_{j1}|\to1\quad \text{as }\quad x\to \infty, \quad j=1, \dots, l. \label{2.10a}
\eer

Now we can state our main result  concerning  the existence and uniqueness of solutions to the BPS equations \eqref{2.7}-\eqref{2.8}  as  the following.

\begin{theorem}\label{th1}
 Consider the BPS system of multiple vortex equation \eqref{2.7}--\eqref{2.8} for \\
 $(q_{11},\dots,q_{l1}, A_{1_1},\dots, A_{l_1}, A_{1_2},\dots, A_{l_2})$
 with prescribed sets of zeros given  by \eqref{2.10} such that $q_{j1}$ have $N_j$ arbitrarily distributed zeros, $j=1, \dots, l$.

 (i) For the problem over a doubly periodic domain $\Omega$, a
 solution exists if and only if
  \be
   \max\limits_{1\le j\le l}\big\{N_j\big\}<\frac{(l+1)|\Omega|}{4\pi}.\label{2.13'}
  \ee

  Furthermore, if there exits a solution, it must be unique.

(ii) For the problem over the full plane $\mathbb{R}^2$ subjected to
the boundary condition  \eqref{2.10a},
there exists a unique solution  up to  gauge transformations such
that the boundary behavior \eqref{2.10a} is reached exponentially
fast
\ber
 \left||q_{il}|^2-1\right|\le  C(\vep)\re^{-(1-\vep)|x|}\quad\mbox{as }|x|\to\infty, \quad i=1, \dots, l\label{e}
\eer
 where $\vep\in (0,1)$ is arbitrarily small, $C(\vep)$ is a positive constant depending on $\vep$.

(iii) In either case, the total vortex fluxes are quantized quantities
given by

 \ber
  \int F_j\ud x=4\pi N_j, \quad\,\, j=1, \dots, l.\label{2.16}
 \eer

\end{theorem}

\begin{remark}
For the first part of the theorem, it can also be proved by a
constrained minimization method developed in  \cite{yang2} and
chapter 4 of  \cite{yang1}, where a more general elliptic system
 was studied. However, in this paper we will prove it
by using a direct minimization method developed in  \cite{liya},
which is more direct and powerful. In fact, following the direct
minimization method used here, we can prove Theorem 1 in
 \cite{yang2}.
\end{remark}

For convenience we complexify the variables. Let
 \[z=x^1+\mathrm{i}x^2,  \quad \tilde{A}_j=A_{j_1}+\mathrm{i}A_{j_2},\quad
 \partial=\frac12(\partial_1-\mathrm{i}\partial_2),\quad \bar{ \partial}=\frac12(\partial_1+\mathrm{i}\partial_2). \]
Hence, noting that $\partial\bar{\partial}=\frac14\Delta$, the BPS equations \eqref{2.7}--\eqref{2.8} are transformed into
 \ber
  \Delta \ln|q_{j1}|^2=|q_{j1}|^2+\sum^l\limits_{i=1}|q_{i1}|^2-(l+1), \,\,\,\,\,j=1,\dots,l,\label{2.9}
 \eer
away from the zeros of $q_{11}, \dots, q_{l1}$.  Then the substitutions
\[u_j=\ln|q_{j1}|^2, \quad\,\, j=1,\dots, l\]
transform the equations \eqref{2.9} into the following nonlinear  elliptic system
 \ber
  \Delta u_j&=&\re^{u_j}+\sum\limits_{i=1}^l\re^{u_i}-(l+1)+4\pi\sum\limits_{s=1}^{N_j}\delta_{p_{j, s}}(x),\quad\,\,j=1,\dots, l, \label{2.11}
  \eer
defined over the entire domain. The boundary condition \eqref{2.10a} now reads
  \be
   u_j\to 0  \quad \text{as}\quad |x|\to \infty, \quad\,\,  j=1,\dots,l. \label{2.12}
  \ee

Throughout this paper we will use the following notations. Let $\mathbf{u}=(u_1,\dots,u_l)^\tau, \mathbf{v}=(v_1,\dots,v_l)^\tau$,
$\mathbf{U}=(\re^{u_1}, \dots, \re^{u_l})^\tau$, and  $A=(a_{ij})$ be the  $l\times l$ matrix
\begin{equation*}
A=\begin{pmatrix}
2 & 1 & 1&\dots & 1\\
1 & 2 & 1&\dots & 1\\
1 &1  & 2 &\dots& 1\\
\vdots&\vdots&\vdots&\ddots&\vdots\\
1 & 1 & 1&\dots & 2
\end{pmatrix}.
\end{equation*}
 For vectors $\mathbf{a}=(a_1, \dots, a_l)^\tau, \,  \mathbf{b}=(b_1, \dots, b_l)^\tau$, we denote
$\mathbf{a}>(\ge)\mathbf{b}$ if $a_i>(\ge)b_i,\, i=1,\dots,l$.

Now the equations \eqref{2.11} can be written in  the vector form
 \be
 \Delta\mathbf{u}=A\mathbf{U}-\mathbf{c}, \label{2.11'}
 \ee
where
\[\mathbf{c}=(c_1,\dots,c_l)\quad \text{with}\quad c_j=(l+1)-4\pi\sum\limits_{s=1}^{N_j}\delta_{p_{j, s}}(x),\quad j=1,\dots,l.\]

To prove Theorem \ref{th1}, it is equivalent to prove the following theorem for the nonlinear  elliptic system \eqref{2.11} or \eqref{2.11'}.
  \begin{theorem}\label{th2}
  Consider the system of nonlinear elliptic equations \eqref{2.11} or \eqref{2.11'}.

  (i)  For the problem over a doubly periodic domain $\Omega$, there exits a
 solution  if and only if the condition
  \be
   \max\limits_{1\le j\le l}\big\{N_j\big\}<\frac{(l+1)|\Omega|}{4\pi}\label{2.13}
  \ee
holds. Moreover, if there is a solution, it must be unique.

 (ii)   For the problem over $\mathbb{R}^2$, there exists a unique solution  satisfying  the boundary condition \eqref{2.12}.
  Moreover, this  boundary condition is achieved exponentially fast,
    \ber
     \sum\limits_{i=1}^l|u_i|^2\le  C(\vep)\re^{-(1-\vep)|x|}\quad\mbox{as }|x|\to\infty, \label{b14'}
     \eer
 where $\vep\in (0,1)$ is arbitrarily small, $C(\vep)$ is a positive constant depending on $\vep$.
  \end{theorem}

  (iii) In both cases, there holds the following quantized integrals
  \ber
   \int \left(\re^{u_j}+\sum\limits_{i=1}^l\re^{u_i}-(l+1)\right)\ud x=-4\pi N_j, \quad j=1, \dots, l. \label{2.13aa}
  \eer

In the  following sections, we just need to  prove Theorem  \ref{th2}.

\section{Proof of existence for doubly periodic case }\label{S:3}
\setcounter{equation}{0}
In this section we prove  Theorem \ref{th2} for the doubly periodic domain case.
 We consider the problem \eqref{2.11} over a doubly periodic domain $\Omega$.

 Let  $u_j^0$  be the solution of (see \cite{aubi} )
 \ber
  \Delta u_j^0&=&4\pi\sum\limits_{s=1}^{N_j}\delta_{p_{j,s}}(x)-\frac{4\pi N_j}{|\Omega|},\quad   x\in\Omega, \quad  j=1,\dots,l.\label{c.2}
 \eer

Set $u_j=u_j^0+v_j,\,j=1,\dots,l$.  The equations \eqref{2.11}
are transformed into
 \ber
 \Delta v_j&=&\re^{u_j^0+v_j}+\sum\limits_{i=1}^l\re^{u_i^0+v_i}-(l+1)+\frac{4\pi N_j}{|\Omega|},\quad  j=1,\dots,l. \label{c.3}
\eer

First we show that the condition \eqref{2.13} is necessary. If $\mathbf{v}=(v_1,\dots,v_l)$ is a solution to the equations
\eqref{c.3}, integrating the equations over $\Omega$ and by a direct computation, we obtain
 \ber
  \ito \re^{u^0_j+v_j}\ud x=|\Omega|-4\pi N_j+\frac{4\pi}{l+1}\sum\limits_{i=1}^lN_i\equiv K_j>0, \quad  j=1,\dots,l.\label{c.1}
 \eer
Then \eqref{2.13} and \eqref{2.13aa} follows.

Next we prove that \eqref{2.13} is also sufficient for the  existence of a solution to \eqref{2.11}. In other words, we will prove that under the condition \eqref{2.13} the
elliptic system \eqref{2.11} admits a unique   solution.

 Let $\mathbf{a}=(a_1,\dots,a_l)^\tau$ with
  \begin{equation*}
\mathbf{a}=\begin{pmatrix}
 (l+1)-\frac{4\pi N_1}{|\Omega|} \\
  \vdots\\
(l+1)-\frac{4\pi N_l}{|\Omega|}
\end{pmatrix}.
\end{equation*}

We can rewrite the equations \eqref{c.3} in a vector form
 \be
  \Delta\mathbf{v}=A\mathbf{U}-\mathbf{a},\label{c.4}
 \ee

To find a variational principle, we need to use the property of the matrix $A$. It is easy to check that the matrix $A$ is positive
definite. Then, by Cholesky decomposition theorem \cite{goor}, we see that the matrix $A$  can be uniquely expressed as a product of a lower triangular  matrix  $L$ and its
transpose, $A=LL^\tau, \,L=(L_{ij})_{l\times l}$.  Indeed, using the iteration scheme in \cite{goor}
 \berr
 && L_{11}=\sqrt{a_{11}}, \quad\,\, L_{j1}=\frac{a_{j1}}{L_{11}},\\
 &&L_{jj}=\sqrt{a_{jj}-\sum\limits_{k=1}^{j-1}L_{jk}^2}=\sqrt{2-\sum\limits_{k=1}^{j-1}L_{jk}^2},\quad\,\,j=1,\dots,l,\\
 &&L_{jk}=\frac{a_{jk}-\sum\limits_{k'=1}^{k-1}L_{jk'}L_{kk'}}{L_{kk}}=\frac{1-\sum\limits_{k'=1}^{k-1}L_{jk'}L_{kk'}}{L_{kk}},
\quad\,\,j=k+1, \dots, l,\quad k=2, \dots, l,
 \eerr
we have
  \berr
 l_{kk}&=&\sqrt{\frac{k+1}{k}}, \quad k=1, \dots, l;\\
  \quad l_{jk}&=&\sqrt{\frac{1}{k(k+1)}}, \quad
  j=2,\dots, l,\quad  k=1,\dots, j-1.
  \eerr
 More explicitly, we obtain
 \be\label{c6a}
L=\begin{pmatrix}
\sqrt2 & 0 & 0&\dots & 0\\
\sqrt{\frac12} &\sqrt{\frac32} & 0&\dots & 0\\
\sqrt{\frac12} &\sqrt{\frac16}  & \sqrt{\frac43}  &\dots& 0\\
\vdots&\vdots&\vdots&\ddots&\vdots\\
\sqrt{\frac12}& \sqrt{\frac16}  & \sqrt{\frac{1}{12}} &\dots &
\sqrt{\frac{l+1}{l}}
\end{pmatrix}.
\ee

 Denote the inverse of $L$ by \[L^{-1}=(\hat{l}_{jk})_{l\times l}.\]
Then, we get
 \berr
  \hat{l}_{kk}=\sqrt{\frac{k}{k+1}}, \quad k=1, \dots,l;\\
  \quad \hat{l}_{jk}=-\sqrt{\frac{1}{{j(j+1)}}}, \quad
  j=2,\dots, l,\quad  k=1,\dots, j-1,
  \eerr
 that is,

 \be\label{c6b}
L^{-1}=\begin{pmatrix}
\sqrt\frac12 & 0 & 0&\dots & 0\\
-\sqrt{\frac16} &\sqrt{\frac23} & 0&\dots & 0\\
-\sqrt{\frac{1}{12}} &-\sqrt{\frac{1}{12}}  & \sqrt{\frac34}  &\dots& 0\\
\vdots&\vdots&\vdots&\ddots&\vdots\\
-\sqrt{\frac{1}{l(l+1)}}& -\sqrt{\frac{1}{l(l+1)}}  &
-\sqrt{\frac{1}{l(l+1)}} &\dots & \sqrt{\frac{l}{l+1}}
\end{pmatrix}.
\ee

 As a result, by $A^{-1}=(LL^\tau)^{-1}=(L^{-1})^\tau L^{-1}$, we have
  \be
A^{-1}=\frac{1}{l+1}\begin{pmatrix}
l & -1 & -1&\dots & -1\\
-1 &l & -1&\dots & -1\\
-1 &-1  & l  &\dots& -1\\
\vdots&\vdots&\vdots&\ddots&\vdots\\
-1& -1& -1&\dots & l
\end{pmatrix}. \label{c5}
\ee

 Let $\mathbf{w}=(w_1,\dots,w_l)^\tau$.  We introduce the transformation
  \be
  \mathbf{w}=L^{-1}\mathbf{v} \quad \text{or}\quad
  \mathbf{v}=L\mathbf{w}\label{c.6},
  \ee
  which can be expressed in the  component form  as (by \eqref{c6a}, \eqref{c6b})
  \ber\left\{\begin{array}{lll}\label{c.6'}
  w_1&=&\frac{1}{\sqrt2}v_1,\\
  w_j&=&-\frac{1}{\sqrt{j(j+1)}}\sum\limits_{i=1}^{j-1}v_i+\sqrt{\frac{j}{j+1}}v_j, \quad j=2, \dots, l,
 \end{array}\right.\eer
 or
 \ber\left\{\begin{array}{lll}\label{c.6''}
  v_1&=&\sqrt2w_1,\\
  v_j&=&\sum\limits_{i=1}^{j-1}\frac{w_i}{\sqrt{i(i+1)}}+\sqrt{\frac{j+1}{j}}w_j, \quad j=2, \dots, l.
 \end{array}\right.\eer

 Let $\mathbf{b}=L^{-1}\mathbf{a}$. Then the system \eqref{c.4}
 becomes
 \be
 \Delta\mathbf{w}=L^\tau\mathbf{U}-\mathbf{b}.\label{c.7}
 \ee
Set $\mathbf{K}=(K_1, \dots, K_l)^\tau$. Hence, by \eqref{c.7} and
\eqref{c.1}, we have
 \be
 \mathbf{K}=|\Omega|(L^\tau)^{-1}\mathbf{b} \quad \text{or}\quad \mathbf{b}=\frac{1}{|\Omega|}L^\tau\mathbf{K}.\label{c.8}
 \ee

 We may  express  \eqref{c.7} in the component form
   \ber
   \Delta w_1&=& \sqrt2\exp\left(u_1^0+\sqrt{2}w_1\right)+\frac{1}{\sqrt2} \sum\limits_{i=2}^l\exp\left(u_i^0+\sum\limits_{k=1}^{i-1}\frac{w_k}{\sqrt{k(k+1)}}+\sqrt{\frac{i+1}{i}}w_i\right)-b_1,\label{c.9}\\
   \Delta  w_j&=& \sqrt{\frac{j+1}{j}}\exp\left(u_j^0+\sum\limits_{k=1}^{j-1}\frac{w_k}{\sqrt{k(k+1)}}+\sqrt{\frac{j+1}{j}}w_j\right)\nm\\
    && +\sqrt{\frac{1}{j(j+1)}}\sum\limits_{i=j+1}^l\exp\left(u_i^0+\sum\limits_{k=1}^{i-1}\frac{w_k}{\sqrt{k(k+1)}}+\sqrt{\frac{i+1}{i}}w_i\right)
    -b_j,\,j=2, \dots,  l-1. \label{c.9aa}\\
    \Delta  w_l&=& \sqrt{\frac{l+1}{l}}\exp\left(u_l^0+\sum\limits_{k=1}^{l-1}\frac{w_k}{\sqrt{k(k+1)}}+\sqrt{\frac{l+1}{l}}w_l\right)-b_l\label{c.9a}
   \eer

It is easy to check that the above system of equations \eqref{c.9}--\eqref{c.9a} are the Euler--Lagrange equations of the functional
   \ber
   I(\mathbf{w})=I(w_1,\dots,w_l)&=&\ito\left\{\frac12\sum\limits_{i=1}^l|\nabla w_i|^2+\exp\left(u_1^0+\sqrt2w_1\right)-b_1w_1\right.\nm\\
   &&\left.+\sum\limits_{i=2}^l\exp\left(u_i^0+\sum\limits_{k=1}^{i-1}\frac{w_k}{\sqrt{k(k+1)}}+\sqrt{\frac{i+1}{i}}w_i\right)-b_iw_i\right\}\ud x.\label{c.10}
   \eer

Let $W^{1,2}(\Omega)$ be the usual Sobolev space of scalar-valued or vector-valued  $\Omega$- periodic $L^2$ functions with their
derivatives also in $L^2(\Omega)$. For $W^{1,2}(\Omega)$ in scalar case, we have the decomposition
\[W^{1,2}(\Omega)=\mathbb{R}\oplus\dot{W}^{1,2}(\Omega)\]
such that  any $w\in W^{1,2}(\Omega)$ can be expressed as
 \be
  w=\underline{w}+\dot{w}, \quad  \underline{w}\in \mathbb{R},\quad \dot{w}\in \dot{W}^{1,2}(\Omega), \quad \ito\dot{w}\ud  x=0.\label{c.11}
 \ee
For any function $w\in\dot{W}^{1,2}(\Omega)$,   there  holds the Trudinger-Moser inequality  \cite{aubi,font}
 \be
 \ito \re^{w}\ud x\le C\exp{\left(\frac{1}{16\pi}\ito|\nabla w|^2\ud x\right)},\label{c.12}
 \ee
which is important for our estimate.

When  $\mathbf{w}\in W^{1,2}(\Omega)$, using the above inequality \eqref{c.12} we see that the functional defined by \eqref{c.10} is a
$C^1$ functional and weakly lower semi-continuous  with respect to the weak topology of $W^{1,2 }(\Omega)$.

For $\mathbf{w}\in W^{1,2}(\Omega)$, applying  \eqref{c.6}, \eqref{c.8} and  the decomposition formula \eqref{c.11}, we have
  \ber
  I(\mathbf{w})&=&\ito\left\{\frac12\sum\limits_{i=1}^l|\nabla \dot{w}_i|^2\right\}\ud x
  +\ito\sum\limits_{i=1}^l\exp(u_0+\dot{v}_i+\underline{v}_i)\ud x
  -\mathbf{K}^\tau\mathbf{\underline{v}}\nonumber\\
  &=&\ito\left\{\frac12\sum\limits_{i=1}^l|\nabla \dot{w}_i|^2\right\}\ud x
  +\ito\sum\limits_{i=1}^l\exp(u_0+\dot{v}_i+\underline{v}_i)\ud x
  -\sum\limits_{i=1}^lK_i\underline{v}_i\label{c.13}
  \eer
Using  Jensen's inequality, we obtain
 \ber
 \ito
\exp(u_i^0+\dot{v}_i+\underline{v}_i)&\ge&|\Omega|\exp\left(\frac{1}{|\Omega|}\ito(u_i^0+\dot{v}_i
  +\underline{v}_i)\ud x\right)\nonumber\\
  &=&|\Omega|\exp\left(\frac{1}{|\Omega|}\ito u_i^0\ud x\right)\re^{\underline{v}_i}\equiv\sigma_i\re^{\underline{v}_i}, \quad  i=1,\dots,l.\label{c.14}
 \eer

Using  the condition \eqref{2.13}, we have $K_i>0,i=1, \dots, l$.
Then, combining \eqref{c.13} and \eqref{c.14}, we have
   \ber
  I(\mathbf{w})-\ito\left\{\frac12\sum\limits_{i=1}^l|\nabla \dot{w}_i|^2\right\}\ud x
   &\ge& \sum\limits_{i=1}^l(\sigma_i\re^{\underline{v}_i}-K_i\underline{v}_i) \nonumber\\
   &\ge& \sum\limits_{i=1}^lK_i\ln\frac{\sigma_i}{K_i}.\label{c.15}
  \eer

 From \eqref{c.15} we can see that the functional $I(\mathbf{w})$ is bounded from
 below in $W^{1,2}(\Omega)$ and the following minimization problem
  \be
   \eta_0\equiv\inf\big\{I(\mathbf{w})|  \mathbf{w} \in W^{1,2}(\Omega)\big\}\label{c.16}
    \ee
is well-defined.

   Let $\Big\{w_1^{(k)},\dots,w_l^{(k)}\Big\}$ be a minimizing sequence of \eqref{c.16}.  It is easy to see  that the
   function $F(t)=\sigma \re^t-\eta t$, where $\sigma, \eta $ are
   positive constants, satisfies the property that  $F(t)\to
   +\infty$ as $t\to \pm\infty$.  Then, we infer from \eqref{c.15} that
    $\Big\{\underline{v}_i^{(k)}\Big\}\, ( i=1,\dots, l)$ are bounded. As a result,
   $\Big\{\underline{w}_i^{(k)}\Big\} \,( i=1,\dots, l)$ are bounded.
   Then, the sequences $\Big\{\underline{w}_i^{(k)}\Big\} \,( i=1,\dots, l)$ admit convergent  subsequences,
   still denoted by $\Big\{\underline{w}_i^{(k)}\Big\} ( i=1,\dots, l)$ for convenience. Then,
    there exist $l$  real numbers $\underline{w}_1^{(\infty)},\dots, \underline{w}_l^{(\infty)}\in
    \mathbb{R}$ such that $\underline{w}_i^{(k)}\to\underline{w}_i^{(\infty)}, i=1, \dots, l$,  as $k\to \infty$.

  In addition, using \eqref{c.15}, we conclude
  that $\Big\{\nabla \dot{w}_i^{(k)}\Big\}\, ( i=1,\dots, l)$  are bounded in $L^2(\Omega)$.
 Therefore, it follows from the  Poincar\'{e} inequality that the sequences $\Big\{\dot{w}_i^{(k)}\Big\} \,( i=1,\dots, l)$ are bounded
  in $W^{1, 2}(\Omega)$.  Consequently,  the sequences $\Big\{\dot{w}_i^{(k)}\Big\} \,( i=1,\dots, l)$ admit weakly convergent subsequences,
  still denoted by $\Big\{\dot{w}_i^{(k)}\Big\}\, ( i=1,\dots, l)$ for convenience. Then, there exist
  $l$ functions $\dot{w}_i^{(\infty)} \in W^{1,2}(\Omega)\, ( i=1,\dots, l)$ such
   that $\dot{w}_i^{(k)}\to\dot{w}_i^{(\infty)}$  weakly in
  $W^{1,2}(\Omega)$ as $k\to\infty \,( i=1,\dots, l)$. Of course, $\dot{w}_i^{(\infty)}\in \dot{W}^{1,
  2}(\Omega)\, ( i=1,\dots, l)$.

    Set
    $w_i^{(\infty)}=\underline{w}_i^{(\infty)}+\dot{w}_i^{(\infty)}\, ( i=1,\dots, l)$, which are all in $W^{1,2}(\Omega)$ naturally. Then, the above
    convergence implies $w_i^{(k)}\to w_i^{(\infty)}\, ( i=1,\dots, l)$ weakly in $W^{1,2}(\Omega)$ as $k\to \infty$.
   Since the functional $I(\mathbf{w})$ is weakly lower semi-continuous in
   $W^{1,2}(\Omega)$, we conclude that $\big(w_1^{(\infty)},\dots, w_l^{(\infty)}\big)$ is a
   solution of the minimization problem \eqref{c.16} and is a critical point of
   $I(\mathbf{w})$. As a critical point of $I(\mathbf{w})$, it satisfies the system \eqref{c.9}.

   Noting that the matrix  $A$ is positive definite, it is easy to check  that $I(\mathbf{w})$ is strictly convex in $W^{1,2}(\Omega)$.
   Then, it has at most one critical point in $W^{1,2}(\Omega)$, which implies the uniqueness of the solution
   to the equations \eqref{c.9}.

\section{Proof of existence for the planar case}
\setcounter{equation}{0}

 In this section we prove Theorem \eqref{th2} for the full plane case. In other words, we study the nonlinear elliptic   system \eqref{2.11} or \eqref{2.11'}  over the full plane with the boundary condition \eqref{2.12}.

 As in  \cite{jata} we introduce the background functions
 \be
 u_j^0=-\sum\limits_{k=1}^{N_j}\ln(1+\mu|x-p_{j,k}|^{-2}),\,\mu>0, \,\, j=1, \dots, l. \label{4.1a}
 \ee
 Then we have
 \ber
  \Delta u_j^0=4\pi\sum\limits_{k=1}^{N_j}\delta_{p_j,k}-g_j,\,\,\,
  g_j=\sum\limits_{k=1}^{N_j}\frac{4\mu}{(\mu+|x-p_{j,k}|^2)^2}, j=1,\dots,l.\label{4.1b}
 \eer
 It is easy to see that
 \be
 g_j\in L(\mathbb{R}^2)\cap L^2(\mathbb{R}^2),\quad \itr g_j\ud x=4\pi N_j, \quad j=1, \dots, l. \label{4.1c}
  \ee

 Let $u_j=v_j+u_j^0, \,\, j=1,\dots,l$,  then the system \eqref{2.11}
 become
   \ber
  \Delta v_j&=&\re^{u^0_j+v_j}+\sum\limits_{i=1}^l\re^{u^0_i+v_i}-(l+1)+g_j,\,\,j=1,\dots, l. \label{4.1}
  \eer

   As in the previous section we use the transformation \eqref{c.6} to change
   \eqref{4.1} into
     \be
     \Delta\mathbf{w}=L^\tau(\mathbf{U}-\mathbf{1})+\mathbf{h} \label{4.2}
   \ee
   or in the component form
   \ber
   \Delta w_1&=& \sqrt2\left[\exp\left(u_1^0+\sqrt{2}w_1\right)-1\right]\nm\\
   &&+\frac{1}{\sqrt2} \sum\limits_{i=2}^l\left[\exp\left(u_i^0+\sum\limits_{k=1}^{i-1}\frac{w_k}{\sqrt{k(k+1)}}+\sqrt{\frac{i+1}{i}}w_i\right)-1\right]+h_1,\label{4.2a}\\
   \Delta  w_j&=& \sqrt{\frac{j+1}{j}}\left[\exp\left(u_j^0+\sum\limits_{k=1}^{j-1}\frac{w_k}{\sqrt{k(k+1)}}+\sqrt{\frac{j+1}{j}}w_j\right)-1\right]\nm\\
    &&+\sqrt{\frac{1}{j(j+1)}}\sum\limits_{i=j+1}^l\left[\exp\left(u_i^0+\sum\limits_{k=1}^{i-1}\frac{w_k}{\sqrt{k(k+1)}}+\sqrt{\frac{i+1}{i}}w_i\right)-1\right]
    +h_j,\nm\\
    &&\quad j=2, \dots,  l-1, \label{4.2bb}\\
    \Delta  w_l&=& \sqrt{\frac{l+1}{l}}\left[\exp\left(u_l^0+\sum\limits_{k=1}^{l-1}\frac{w_k}{\sqrt{k(k+1)}}+\sqrt{\frac{l+1}{l}}w_l\right)-1\right]+h_l,\label{4.2b}
    \eer
 where \[\mathbf{1}=(1, \dots, 1)^\tau, \qquad \mathbf{h}=(h_1,\dots,h_l)^\tau=L^{-1}\mathbf{g}, \qquad\mathbf{g}=(g_1,\dots, g_l)^\tau .\]
 It is easy to check that \eqref{4.2} are the Euler--Lagrange
 equations of the following functional
  \ber
    I(\mathbf{w})&=&\int_{\mathbb{R}^2}\left\{\frac12\sum\limits_{i=1}^l|\nabla w_i|^2+\sum\limits_{i=1}^lh_iw_i+\exp\left(u_1^0+\sqrt2w_1\right)-\exp(u_1^0)-\sqrt2w_1\right.\nonumber\\
  &&\left.+\sum\limits_{i=2}^l\left[\exp\left(u_i^0+\sum\limits_{k=1}^{i-1}\frac{w_k}{\sqrt{k(k+1)}}+\sqrt{\frac{i+1}{i}}w_i\right)-\exp(u_i^0)\right.\right.\nm\\
  &&\left.\left.- \sum\limits_{k=1}^{i-1}\frac{w_k}{\sqrt{k(k+1)}}-\sqrt{\frac{i+1}{i}}w_i\right]\right\}\ud x.\label{4.3}
  \eer

To proceed further, it is convenient to rewrite the functional $I$ as the following form
 \berr
  I(\mathbf{w})&=&\int_{\mathbb{R}^2}\left\{ \frac12\sum\limits_{i=1}^l|\nabla w_i|^2\ud x+ \re^{u_1^0}\left[\re^{\sqrt2w_1}-1-\sqrt2w_1\right]\right.\nm\\
  &&+\sum\limits_{i=2}^l\re^{u_i^0}\left[\exp\left(\sum\limits_{k=1}^{i-1}\frac{w_k}{\sqrt{k(k+1)}}+\sqrt{\frac{i+1}{i}}w_i\right)-1-\sum\limits_{k=1}^{i-1}\frac{w_k}{\sqrt{k(k+1)}}-\sqrt{\frac{i+1}{i}}w_i\right]\nm\\
  &&+\sqrt2\left(\re^{u_1^0}-1+\frac{1}{\sqrt2}h_1\right)w_1+\sqrt{\frac{l+1}{l}}\left(\re^{u_l^0}-1\right)w_l\nm\\
  &&\left.+\sum\limits_{i=2}^{l-1}\left[\sqrt{\frac{i+1}{i}}\left(\re^{u_i^0}-1\right)+\frac{1}{\sqrt{i(i+1)}}\sum\limits_{k=i+1}^l\left(\re^{u_k^0}-1\right)+h_i\right]w_i\right\}\ud x
 \eerr

Then we obtain
 \ber
(DI(\mathbf{w}))(\mathbf{w})&=&\int_{\mathbb{R}^2}\left\{\sum\limits_{i=1}^l|\nabla w_i|^2\ud x \ud x+\left[\sqrt2(\exp(u_1^0+\sqrt2w_1)-1)\right.\right.\nm\\
 &&\left.+\frac{1}{\sqrt2}\sum\limits_{i=2}^l\left(\exp\left(u_i^0+\sum\limits_{k=1}^{i-1}\frac{w_k}{\sqrt{k(k+1)}}+\sqrt{\frac{i+1}{i}}w_i\right)-1\right)\right]w_1\nm\\
 &&+\sum\limits_{j=2}^{l-1}\left[\sqrt{\frac{j+1}{j}}\left[\exp\left(u_j^0+\sum\limits_{k=1}^{j-1}\frac{w_k}{\sqrt{k(k+1)}}+\sqrt{\frac{j+1}{j}}w_j\right)-1\right]\right.\nm\\
 &&\left.+\frac{1}{\sqrt{j(j+1)}}\sum\limits_{i=j+1}^l\left[\exp\left(u_i^0+\sum\limits_{k=1}^{i-1}\frac{w_k}{\sqrt{k(k+1)}}+\sqrt{\frac{i+1}{i}}w_i\right)-1\right]\right]w_j\nm\\
 &&+\sqrt{\frac{l+1}{l}}\left[\exp\left(u_l^0+\sum\limits_{k=1}^{l-1}\frac{w_k}{\sqrt{k(k+1)}}+\sqrt{\frac{l+1}{l}}w_l\right)-1\right]w_l+\sum\limits_{j=1}^lh_jw_j\nm\\
 &&=\int_{\mathbb{R}^2}\sum\limits_{i=1}^l|\nabla w_i|^2\ud x \ud x+\left[\exp\left(u_1^0+\sqrt2w_1\right)-1+\tilde{h}_1\right]\sqrt2w_1\nm\\
 &&+\sum\limits_{i=2}^l\left[\exp\left(u_i^0+\sum\limits_{k=1}^{i-1}\frac{w_k}{\sqrt{k(k+1)}}+\sqrt{\frac{i+1}{i}}w_i\right)-1+\tilde{h}_i\right]\nm\\
 &&\left.\times\left(\sum\limits_{k=1}^{i-1}\frac{w_k}{\sqrt{k(k+1)}}+\sqrt{\frac{i+1}{i}}w_i\right)\right\}\ud x,\label{d7}
 \eer
where   \[ \tilde{\mathbf{h}}\equiv(\tilde{h}_1, \dots, \tilde{h}_l)^\tau=(L^{-1})^\tau \mathbf{h}=(L^{-1})^\tau L^{-1}\mathbf{g}=A^{-1}\mathbf{g},\]\label{d8}
or in the component form (using \eqref{c5})
\be
\tilde{h}_i=\frac{1}{l+1}\left(lg_i-\sum\limits_{j\neq i}^lg_j\right),\quad i=1, \dots, l.
\ee

Noting the transformation \eqref{c.6}, we easily see that
 \ber
   c_1\sum\limits_{i=1}^mv_i^2\le\sum\limits_{i=1}^mw_i^2\le c_2\sum\limits_{i=1}^mv_i^2, \quad
   c_1\sum\limits_{i=1}^m|\nabla v_i|^2\le\sum\limits_{i=1}^m|\nabla w_i|^2\le c_2\sum\limits_{i=1}^m|\nabla v_i|^2 \label{d9}
  \eer
 holds for some positive constants $c_1$ and  $c_1$.
Therefore, from \eqref{c.6''}, \eqref{d7} and \eqref{d9}, we can obtain
 \ber
 (DI(\mathbf{w}))(\mathbf{w}) \ge C_1\sum\limits_{i=1}^l\itr|\nabla v_i|^2\ud x+\sum\limits_{i=1}^l\itr\left(\re^{u_0^i+v_i}-1+\tilde{h}_i\right)v_i\ud x,  \label{d10}
  \eer
 where   and  in the sequel  we use  $C_i$ to denote a generic positive constant.

In what follows we estimate   the second term on the right hand side of \eqref{d10}. To this end we use the approach developed in  \cite{jata}.
 It is sufficient to deal with a general term of  the following  form
  \berr
  G(v)=\itr\left(\re^{u^0+v}-1+\tilde{h}\right)v\ud x,
  \eerr
  where we use $u^0$, $v$, and  $\tilde{h}$ to denote $u_i^0$'s, $v$'s and $\tilde{h}_i$'s, respectively.
For convenience, we decompose  $G(v)$  as  $G(v)=G(v_+)+G(-v_-)$
where $v_+=\max\{v, \,0\}, v_-=\max\{-v, \,0\}$.

Using the inequality  $\re^t-1\ge t, \,t\in \mathbb{R}$ and the fact $u^0,\, \tilde{h}\in L^2(\mathbb{R}^2)$,  we obtain
 \ber
 G(v_+)\ge\itr(u^0+v_++\tilde{h})v_+\ud x\ge \frac12\itr v_+^2\ud x-C_2.\label{d11}
 \eer

 In view of the inequality $1-\re^{-t}\ge\frac{t}{1+t},\, t\ge0$,  $\re^{u^0}-1$, $\tilde{h}\in L^2(\mathbb{R}^2)$,  we estimate  $G(-v_-)$ as follows
 \ber
   G(-v_-)&=&\itr\left(1-\re^{u^0-v_-}-\tilde{h}\right)v_-\ud x\nm\\
   &=&\itr\left(1-\tilde{h}-\re^{u^0}+\re^{u^0}\left[1-\re^{-v_-}\right]\right)v_-\ud x\nm\\
   &\ge&\itr\left(1-\tilde{h}-\re^{u^0}+\re^{u^0}\frac{v_-}{1+v_-}\right)v_-\ud x\nm\\
   &=&\itr\left([1+v_-]\left[1-\tilde{h}-\re^{u^0}\right]+\re^{u^0}v_-\right)\frac{v_-}{1+v_-}\ud x\nm\\
   &=&\itr\left(1-\tilde{h}\right)\frac{v_-^2}{1+v_-}\ud x+\itr \left(1-\re^{u^0}-\tilde{h}\right)\frac{v_-}{1+v_-}\ud x \nm\\
   &\ge&\frac12\itr\frac{v_-^2}{1+v_-}\ud x+\itr \left(1-\re^{u^0}-\tilde{h}\right)\frac{v_-}{1+v_-}\ud x \nm\\
   &\ge&\frac14 \itr\frac{v_-^2}{(1+v_-)^2}\ud x-C_3,\label{d12}
 \eer
 where we have used the fact $\tilde{h}\le \frac12$, assured by taking $\mu$ sufficiently large.

Therefore from \eqref{d11} and \eqref{d12} we conclude that
 \be
  G(v)\ge \frac14\itr\frac{v^2}{(1+|v|)^2}\ud x-C_4. \label{d13}
 \ee

  By the following   standard interpolation inequality over $\wot$:
 \be
 \itr v^4\ud x\le 2\itr v^2\ud x\itr |\nabla v|^2\ud x, \quad \forall\, v\in \wot, \label{d14}
 \ee
  we have
  \berr
   &&\left(\itr|v|^2\ud x\right)^2\nm\\
   &&=\left(\itr\frac{|v|}{1+|v|}(1+|v|)|v|\ud  x\right)^2\nm\\
    &&\le\itr\frac{|v|^2}{(1+|v|)^2}\ud x\itr\big(|v|+|v|^2\big)^2\ud x\nm\\
    &&\le4\itr\frac{|v|^2}{(1+|v|)^2}\ud x\itr|v|^2\ud x\left(\itr|\nabla v|^2\,\ud x+1\right)\nm\\
    &&\le\frac12 \left(\itr|v|^2\ud x\right)^2+C\left(\left[\frac{|v|^2}{(1+|v|)^2}\ud x\right]^4 +\left[\itr|\nabla v|^2\ud x\right]^4+1\right),
  \eerr
  which  implies
   \ber
   \|v\|_2&\le& C_5\left(\itr \frac{|v|^2}{(1+|v|)^2}\ud x+\itr|\nabla v|^2\ud x+1\right), \label{d15}
   \eer
where and in the sequel we use $\|\cdot\|_p$ to denote the norm of the space $L^p(\mathbb{R}^2)$.

From \eqref{d10} and \eqref{d13}, we obtain
  \be
   (DI(\mathbf{w}))(\mathbf{w})\ge C_6\sum\limits_{j=1}^m\itr\left(|\nabla v_j|^2+\frac{v^2_j}{(1+|v_j|)^2}\right)\ud x-C_7 .\label{d16}
  \ee
Then it follows from \eqref{d9}, \eqref{d10} and \eqref{d16} that
  \ber
  (DI(\mathbf{w}))(\mathbf{w})\ge  C_8\sum\limits_{j=1}^m\|w_j\|_{\wot}-C_9.\label{d17}
  \eer

Now we  show that the functional $I$ has a critical point.  Using  \eqref{d17},   we can choose $R>0$ such that
   \be
    \inf\{DI(\mathbf{w})(\mathbf{w}) \,|\,\|\mathbf{w}\|_{\wot}=R \}\ge 1.
   \ee
Noting that  functional $I$ is weakly  lower semi-continuous on $\wot$, we see that the minimization problem
  \be
  \eta_0\equiv\inf\left\{I(\mathbf{w})|\, \|\mathbf{w}\|_{\wot}\le  R\right\} \label{d19}
  \ee
admits a solution, say, $\hat{\mathbf{w}}$.  We may prove that it must be an interior point.  Otherwise, we assume that $\|\hat{\mathbf{w}}\|_{\wot}= R$.  Therefore
 \berr
  \lim\limits_{t\to0}\frac{I((1-t)\hat{\mathbf{w}})-I(\hat{\mathbf{w}})}{t}
  = \frac{\ud }{\ud t}I((1-t)\tilde{\mathbf{w}})\big|_{t=0}=-(DI(\hat{\mathbf{w}}))(\hat{\mathbf{w}})\le -1
 \eerr
Hence, if $t>0$ is sufficiently small, set  $\hat{\mathbf{w}}^t=(1-t)\hat{\mathbf{w}}$, we see that
 \berr
I(\hat{\mathbf{w}}^t)<I(\hat{\mathbf{w}})=\eta_0, \quad \|\hat{\mathbf{w}}^t\|_{\wot}=(1-t)R<R,
\eerr
which contradicts the definition of $\eta_0$. Hence, $\hat{\mathbf{w}}$ must be an interior critical  point for the
problem \eqref{d19}. As a result, it is a critical point of the functional $I$. Since the functional $I$  is strictly convex, this critical point must be unique.

Now  we  investigate the asymptotic behavior of the solution established  above.  Since   $\mathbf{w}\in \wot$,  using the well-known inequality
  \berr
   \|\re^v-1\|_2^2\le d_1\exp(d_2\|v\|_{\wot}^2),\quad \forall \,v\in \wot,
  \eerr
where $d_1, d_2$ are some positive constants, we see that the right-hand sides of the equations \eqref{4.2a}--\eqref{4.2b} all belong to $L^2(\mathbb{R}^2)$.
By the  standard elliptic $L^2$-estimates, we obtain that  $w_i\in W^{2,2}(\mathbb{R}^2)$, which implies $w_i\to 0$ as
$|x|\to \infty$, $i=1,\dots, l$. Using the transformation \eqref{c.6''}, we conclude  that $v_i\to 0$ as $|x|\to \infty$, which gives the desired
boundary condition $u_i\to 0$ as $|x|\to \infty, \, i=1, \dots, l$.

  Now  we prove that $|\nabla w_i|\to0$ as $|x|\to\infty$, $i=1,\dots,l$.  We rewrite a  typical term of the right hand sides of \eqref{4.2a}-\eqref{4.2b} as
 \ber
 &&\sqrt{\frac{j+1}{j}}\left[\exp\left(u_j^0+\sum\limits_{k=1}^{j-1}\frac{w_k}{\sqrt{k(k+1)}}+\sqrt{\frac{j+1}{j}}w_j\right)-1\right]\nm\\
&&=\sqrt{\frac{j+1}{j}}\left[\left(\re^{u_j^0}-1\right)\exp\left(\sum\limits_{k=1}^{j-1}\frac{w_k}{\sqrt{k(k+1)}}+\sqrt{\frac{j+1}{j}}w_j\right)\right.\nm\\
&&\left.+\exp\left(\sum\limits_{k=1}^{j-1}\frac{w_k}{\sqrt{k(k+1)}}+\sqrt{\frac{j+1}{j}}w_j\right)-1\right]
\eer
which lies in  $L^p(\mathbb{R}^2)$ for any $p>2$ due to the embedding $\wot\subset L^p(\mathbb{R}^2)$ and the definition of
$u^0_j$.  Then we see that  all the right-hand-side terms of \eqref{4.2a}-\eqref{4.2b} belong to $L^p(\mathbb{R}^2)$, for any $p>2$.
Using the elliptic $L^p$-estimates, we see that $w_i\in W^{2,p}(\mathbb{R}^2)$ for any $p>2$, $i=1,\dots, l$.  As a result,
 $|\nabla w_i|\to 0$ as $|x|\to \infty$, $i=1,\dots l$. In other words,  $|\nabla u_i|\to 0$ as $|x|\to \infty$, $i=1,\dots l$.

Next we  establish the exponential decay rate of the solutions at infinity. To do this, we consider the equations
\eqref{2.11} or \eqref{2.11'} over an exterior domain  \be D_R=\left\{x\in \mathbb{R}^2| \quad |x|>R\right\},
\ee
where $R>0$ satisfies
\be
R>\max\big\{ |p_{i,s}|\, | \, i=1,\dots,l,\,s=1,\dots,N_i\big\}.
\ee

For convenience, we consider the system of equations \eqref{2.11'} over $D_R$. We may rewrite (\ref{2.11'}) in $D_R$ as
\be
\Delta {\bf u}=A({\bf U}-{\bf 1})= A{\bf u}+A({\bf U}-{\bf 1}-{\bf u}). \label{d23}
\ee

  Since the matrix $A$  in \eqref{2.11'} is positive definite and its
 eigenvalues  are $\lambda_1=l+1, \lambda_2=\lambda_3=\cdots=\lambda_l=1$, which can be checked easily.
Then there exists an orthogonal matrix $O$ such that
\be
 O^\tau AO={\rm diag}\{l+1, 1, \dots, 1\}.
\ee

Now apply $O^\tau$ in (\ref{d23}) and set
 \[
 \tilde{\mathbf{u}}=O^\tau{\bf u}.
 \]
Then we have
 \ber
 \Delta \tilde{\mathbf{u}}={\rm diag}\{l+1, 1, \dots, 1\}\tilde{\mathbf{u}}+O^\tau ({\bf U}-{\bf 1}-{\bf u}). \label{d25}
 \eer
 Noting that  ${\bf U}\to {\bf 1}$ as $|x|\to\infty$, we have ${\bf U}-{\bf 1}=E(x){\bf u}$, where $E(x)$ is an
$l\times l$ diagonal matrix so that $E(x)\to I_l$ (the $l\times l$ identity matrix) as $|x|\to\infty$.  Then we can  rewrite
 \eqref{d25} as
\be
\Delta \tilde{\mathbf{u}}={\rm diag}\{l+1, 1, \dots, 1\}\tilde{\mathbf{u}} +Z(x)\tilde{\bf u},\label{d26}
\ee
where $Z(x)$ is an $l\times l$ matrix which vanishes at infinity.   Hence from (\ref{d26}) we see that
  \ber
  \Delta|\tilde{\mathbf{u}}|^2\ge 2\tilde{\mathbf{u}}^\tau\Delta
  \tilde{\mathbf{u}}\ge |\tilde{\mathbf{u}}|^2- b(x)|\tilde{\mathbf{u}}|^2,
  \eer
 where $b(x)\to 0$ as $|x|\to \infty$.

Therefore, for any $\vep\in (0,1)$, we can take  a suitably large $R_\vep\geq R$ such that
 \be \label{d27}
 \Delta|\tilde{\mathbf{u}}|^2\ge\left(1- \frac{\vep}{2}\right)|\tilde{\mathbf{u}}|^2, \quad x\in D_{R_\vep}.
 \ee

 Taking a  comparison function, say $\eta$, of the form
 \be\label{d28}
 \eta = C\re^{-\sigma|x|},\quad |x|>0,\quad C,\sigma\in\mathbb{R},\quad C,\sigma>0
 \ee
we have  $\Delta\eta=\sigma^2\eta-\frac\sigma{|x|}\eta$.   Hence,  by  (\ref{d27})  we  obtain
\be
\Delta\left(|\tilde{\bf u}|^2-\eta\right)\geq\left(1-\frac\vep2\right) |\tilde{\bf u}|^2-\sigma^2\eta,\quad |x|\geq R_\vep.
\ee
We take the obvious choice $\sigma^2=\left(1-\frac\vep2\right)\lambda\lambda_0$ which gives us
$\Delta(|\tilde{\bf u}|^2-\eta)\geq\sigma^2(|\tilde{\bf u}|^2-\eta)$, $|x|\geq R_\vep$.
Taking $C$ in (\ref{d28}) large  such  that $|\tilde{\bf u}|^2-\eta\leq 0$ for $|x|= R_\vep$,
  using the fact that $|\tilde{\mathbf{u}}|\to 0$ as $|x|\to \infty$  and
the maximum principle,   we conclude that $|\tilde{\bf u}|^2\leq\eta$ for $|x|\geq R_\vep$. Noting  $(1-\vep)^2<\left(1-\frac\vep2\right)$
for any $\vep\in(0,1)$, we get the precise exponential decay  estimate
\be
|\tilde{\mathbf{u}}|^2\le C(\vep)\re^{-(1-\vep)|x|},\quad |x|\geq R_\vep.
\ee

 Thus we get the desired exponential decay rate \eqref{b14'}.

At last, we aim to prove the quantized integrals. To this end, we need to  establish the exponential decay rate for the derivatives.

Let $\partial$ denote any of the two derivatives $\partial_1$ and  $\partial_2$.  Define
 \be
{\mathbf{v}}=(\partial u_1, \dots, \partial u_m)^\tau,\quad M={\rm diag}\{\re^{u_1}-1, \dots, \re^{u_l}-1\}.
 \ee
Then differentiating  \eqref{2.11'} in $D_R$, we have
  \be
   \Delta {\mathbf{v}}= A{\mathbf{v}}+A M{\mathbf{v}}.\label{d33}
  \ee
Let $O$ be as before and set
\be
{\mathbf{v}}=O \tilde{\mathbf{v}}.
\ee
  Then by \eqref{d33},   we have
  \ber
   \Delta \tilde{\mathbf v}={\rm diag}\{l+1, 1, \dots, 1\}\tilde{\mathbf{v}}+O^\tau A M O\tilde{\mathbf v}.\label{d34}
  \eer
  Since  $\re^{u_i}-1\to 0$ as $|x|\to \infty$, $i=1, \dots,l$, we may rewrite  \eqref{d34} as
    \ber
   \Delta \tilde{\mathbf v}={\rm diag}\{l+1, 1, \dots, 1\}\tilde{\mathbf{v}}+\tilde{Z}(x)\tilde{\mathbf v}, \label{d34}
  \eer
  where $\tilde{Z}(x)$ is an $l\times l$ matrix vanishing at infinity.
  As previously, from \eqref{d34} we have
 \ber
  \Delta|\tilde{\mathbf{v}}|^2\ge 2\tilde{\mathbf{v}}^\tau\Delta \tilde{\mathbf{v}}\ge |\tilde{\mathbf{v}}|^2- \tilde{b}(x)|\tilde{\mathbf{v}}|^2,
  \eer
 where $\tilde{b}(x)\to 0$ as $|x|\to \infty$.

Similarly, we may infer that,  for any $\vep\in(0,1)$,  there is a positive constant $C(\vep)>0$, such that
\[
|\mathbf{\tilde{v}}|^2\le C(\vep)\re^{-(1-\vep)|x|},
\]
when $|x|$ is sufficiently large.
 Then we obtain the  exponential decay rate near infinity
 \be
  \sum\limits_{i=1}^m|\nabla u_i(x)|^2\le C(\vep)\re^{-(1-\vep)|x|}.\label{d40}
 \ee

\medskip

Now we can  calculate  the quantized integrals \eqref{2.13aa} stated in Theorem \ref{th2} for the planar case.

Using \eqref{4.1a}, \eqref{4.1b}, and  the exponential decay property  of $|\nabla u_i|$'s in \eqref{d40}, we conclude that $|\nabla  v_i|$'s vanish
at infinity at least at the rate $|x|^{-3}$.  Then it follows from the divergence theorem that
\be \label{d42}
\int_{\mathbb{R}^2}\Delta v_i\,\ud x=0,\quad i=1,\dots,l.
\ee
Thus, by integrating the equations \eqref{4.1} over $\mathbb{R}^2$, and applying \eqref{4.1c} and (\ref{d42}), we  obtain the desired results stated in \eqref{2.13aa}.

\section{Conclusions}

We have carried out a rigorous analysis of the BPS equations derived from  the theory of multi-intersections of $l \,(l\ge2)$ D4-branes and 1 D4-brane.
The BPS equations are investigated  in two situations.   In  the first situation the equations are studied over a doubly periodic domain. In the second
situation  the equations are studied over the full plane. Via the direct minimization method we establish a  sharp existence and
uniqueness theorem  for multiple vortex solutions of the BPS equations. We find  an explicit necessary and sufficient condition
for the existence of a unique solution for the doubly periodic domain case. By the obtained vortex solutions we can interpret them as D0-branes on the intersections.

\end{document}